\documentclass[twocolumn,showpacs,preprintnumbers,amsmath,amssymb,superscriptaddress]{revtex4}
\usepackage{graphicx}% Include Fig. files
\usepackage{dcolumn}% Align table columns on decimal point
\usepackage{bm}% bold math
\usepackage{graphicx}
\usepackage{xspace}
\usepackage{multirow}
\usepackage{natbib}
\usepackage{color}

\renewcommand{\vec}{\mathbf}

\begin{document}

\preprint{}

\title{Strong nodeless pairing on separate electron Fermi surface sheets in (Tl, K)Fe$_{1.78}$Se$_2$ probed by ARPES}

\author{X.-P. Wang}
\affiliation{Beijing National Laboratory for Condensed Matter Physics, and Institute of Physics, Chinese Academy of Sciences, Beijing 100190, China}
\author{T. Qian}
\affiliation{Beijing National Laboratory for Condensed Matter Physics, and Institute of Physics, Chinese Academy of Sciences, Beijing 100190, China}
\author{P. Richard}
\affiliation{Beijing National Laboratory for Condensed Matter Physics, and Institute of Physics, Chinese Academy of Sciences, Beijing 100190, China}
\author{P. Zhang}
\affiliation{Beijing National Laboratory for Condensed Matter Physics, and Institute of Physics, Chinese Academy of Sciences, Beijing 100190, China}
\author{J. Dong}
\affiliation{Beijing National Laboratory for Condensed Matter Physics, and Institute of Physics, Chinese Academy of Sciences, Beijing 100190, China}
\author{H.-D. Wang}
\affiliation{Department of Physics, Zhejiang University, Hangzhou 310027, China}
\author{C.-H. Dong}
\affiliation{Department of Physics, Zhejiang University, Hangzhou 310027, China}
\author{M.-H. Fang}
\affiliation{Department of Physics, Zhejiang University, Hangzhou 310027, China}
\author{H. Ding}\email{dingh@iphy.ac.cn}
\affiliation{Beijing National Laboratory for Condensed Matter Physics, and Institute of Physics, Chinese Academy of Sciences, Beijing 100190, China}

\date{\today}% It is always \today, today,

\begin{abstract}
We performed a high-resolution angle-resolved photoemission spectroscopy study of the Tl$_{0.63}$K$_{0.37}$Fe$_{1.78}$Se$_2$ superconductor ($T_c=29$ K). We show the existence of two electronlike bands at the M$(\pi, 0)$ point which cross the Fermi level at similar Fermi wave vectors to form nearly circular electronlike Fermi surface pockets. We observe a nearly isotropic $\sim$ 8.5 meV superconducting gap ($\Delta/k_BT_c\sim 7$) on these Fermi surfaces. Our analysis of the band structure around the Brillouin zone centre reveals two additional electronlike Fermi surfaces: a very small one and a larger one with $k_F$ comparable to the FS pockets at M. Interestingly, a SC gap with a magnitude of $\sim$ 8 meV also develops along the latter FS. Our observations are consistent with the $s$-wave strong coupling scenario. 
\end{abstract}

%\verb+\pacs{#1}+ command.

%Use showkeys class option if keyword
\maketitle

The amplitude and symmetry of the superconducting (SC) gap of a material are determined by its band structure, its Fermi surface (FS) topology and the pairing mechanism itself. The experimental observation of enhanced gap amplitude on holelike and electronlike FS pockets quasi-nested by the antiferromagnetic (AF) wave vector in iron-based superconductors \cite{Ding_EPL, L_Zhao, Nakayama_EPL2009, Terashima_PNAS2009, Nakayama_PRL2010} has been widely considered as suggestive of the importance of AF interband scattering in these materials. In particular, the quasi-nesting model is consistent with the strong suppression of superconductivity in heavily hole-doped \cite{Sato_PRL2009} and heavily electron-doped \cite{Sekiba_NJP2009} BaFe$_2$As$_2$ compounds, for which the FS quasi-nesting conditions vanish. Recently, this model faced a serious challenge with the discovery of superconductivity above 30 K in heavily electron-doped K$_{0.8}$Fe$_{2-x}$Se$_2$ and (Tl,K)Fe$_{2-x}$Se$_2$ \cite{JG_Guo_PRB2010, MH_Fang1012}. Indeed, previous angle-resolved photoemission spectroscopy (ARPES) measurements revealed only electronlike FS pockets \cite{Y_Zhang_arxiv2010, Qian_KFeSe2010}. 

In this letter, we report high-energy resolution ARPES measurements on the Tl$_{0.63}$K$_{0.37}$Fe$_{1.78}$Se$_2$ superconductor ($T_c=29$ K). We observed two electronlike M$(\pi, 0)$-centred FS pockets that develop a nearly isotropic SC gap below $T_c$ with a magnitude of $\sim$ 8.5 meV, leading to a 2$\Delta$/$k_BT_c$ of $\sim$ 7. In addition, a weak electronlike FS pocket with a similar size and a tiny electronlike pocket are also observed at the $\Gamma (0,0)$ point. The former one also exhibits a SC gap size of about 8 meV. In addition, a high-energy ($\sim$ 0.8 eV) incoherent peak undergoes a significant energy shift of $\sim$ 100 meV through the metal-nonmetal crossover around 70K, while the low-energy valence band shows little change. We discuss the possible implications of the SC gap symmetry and the FS topology for the SC pairing mechanism in this unusual iron-based superconductor .

\begin{figure*}
\includegraphics[width=16cm]{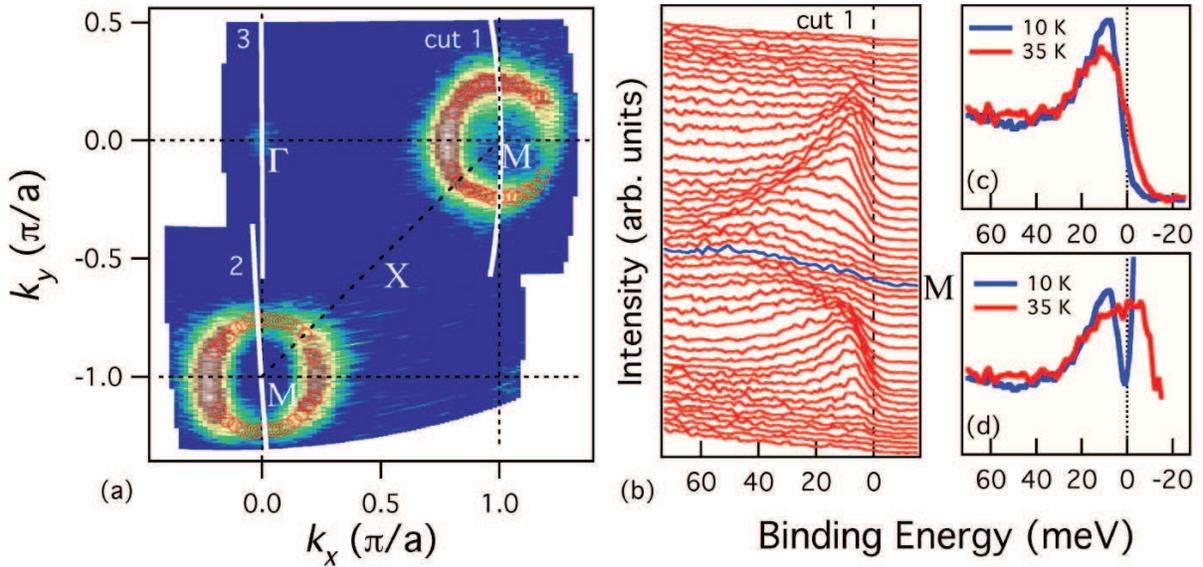}
\caption{\label{Fig_map}(Colour on-line) (a) Momentum-resolved photoemission intensity mapping of Tl$_{0.63}$K$_{0.37}$Fe$_{1.78}$Se$_2$ recorded in the normal state (35 K) and integrated over a 10 meV window centred at $E_F$. The small red circles indicate the FS obtained from the momentum distribution curve (MDC) peak position at $E_F$. (b) EDCs recorded along cut1 from panel (a). The blue line indicate the spectrum at the M point. (c) EDC at $k_F$ recorded at 10 and 35 K. (d) Same as (c) but after division by the Fermi-Dirac function.}
\end{figure*}

Single crystals of Tl$_{0.63}$K$_{0.37}$Fe$_{1.78}$Se$_2$ ($T_c^{onset}=29.1$ K; $T_c^{mid}=28.6$ K; $T_c^{zero}=27.5$ K) were grown by the Bridgeman method \cite{MH_Fang1012}. The precise composition was determined using an Energy Dispersive X-ray Spectrometer (EDXS). The lattice parameters $a=3.85$ \AA\xspace and $c=14.05$ \AA\xspace were obtained by fitting XRD data. We performed ARPES measurements at the Institute of Physics, Chinese Academy of Sciences using the He I$\alpha$ resonance line (h$\nu$ = 21.218 eV). The angular resolution was set to 0.2$^{\circ}$ while the energy resolution was set to 4-7 meV for high-resolution measurements. Samples with a typical size of $\sim 2 \times 2$ mm$^2$ were cleaved \emph{in situ} and measured between 10 and 35 K in a working vacuum better than 5$\times$10$^{-11}$. The Fermi energy ($E_F$) of the samples was referenced to that of a gold film evaporated onto the sample holder. For convenience, we describe all the results using the 1 Fe site/unit (or unfolded) cell notation.

In Fig. \ref{Fig_map}(a), we show a momentum-resolved photoemission intensity mapping of Tl$_{0.63}$K$_{0.37}$Fe$_{1.78}$Se$_2$ recorded in the normal state (35 K) and integrated over a 10 meV window centred at the Fermi level ($E_F$). As with other A$_{x}$Fe$_2$Se$_2$ materials with similar $T_c$ values \cite{Qian_KFeSe2010, Y_Zhang_arxiv2010}, the dominant feature of the mapping is an almost circular electronlike FS pocket centred at the M point. We note that the intensity patterns around M1$(\pi,0)$ and M2$(0,\pi)$ are different due to different matrix elements resulting from different $\vec{A}\cdot\vec{p}$ configurations, where $\vec{A}$ is the potential vector associated with the incoming light. In order to extract the SC gap size precisely along the FS and to investigate its symmetry, we performed high-resolution measurements along cuts crossing the Fermi surface. In Fig. \ref{Fig_map}(b), we display a series of energy distribution curves (EDCs) along cut1 given in Fig. \ref{Fig_map}(a). The EDCs show a peak dispersing towards $E_F$ that starts bending back at the Fermi wave vector ($k_F$), a hallmark of Bogoliubov dispersion in the SC state. By comparing SC-state and normal-state EDCs measured at the same $k_F$, as shown in Fig. \ref{Fig_map}(c), one sees that the leading edge of the SC-state EDC shifts away from $E_F$. After the Fermi-Dirac function is divided out from the spectra [Fig. \ref{Fig_map}(d)], the normal-state EDC exhibits a peak at $E_F$, while the SC-state EDC remains gapped.

\begin{figure}
\includegraphics[width=8.6cm]{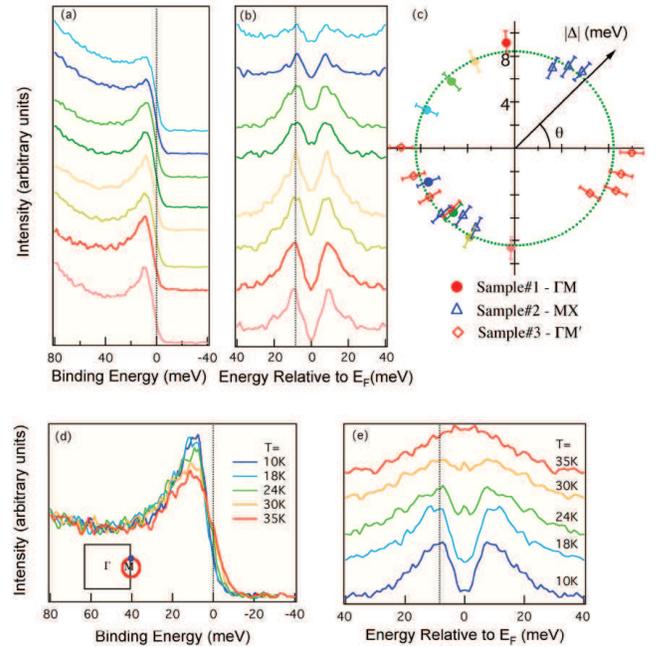}
\caption{\label{Fig_gap}(Colour on-line) (a) Spectra recorded in the SC state (10 K) at different location around the M-centred electron like FS. The colour of the spectra correspond to their angular locations, which are given in panel (c) (sample 1; plain symbols). (b) Corresponding symmetrized EDCs. (c) Polar distribution of the SC gap size along the FS for 3 different samples. (d) Temperature dependence of EDC spectra at $k_F$. The inset show the momentum location of the EDC. (e) Corresponding symmetrized EDCs.}
\end{figure}

The raw energy distribution curves (EDCs) at $k_F$ are given in Fig. \ref{Fig_gap}(a). Following a standard procedure in ARPES, we take advantage of the particle-hole symmetry at $k_F$ by symmetrizing the spectra to approximately remove the Fermi-Dirac function. The symmetrized EDCs are shown in Fig. \ref{Fig_gap}(b). The SC gap size $\Delta$ is approximated by half the distance between the two peaks. Obviously, the peak position does not vary much with momentum within experimental uncertainties. We report the gap size extracted for different samples mounted with different orientations in Fig. \ref{Fig_gap}(c). We find that the SC gap size averages at $\sim$ 8.5 meV with little room for anisotropy and even less for nodes, at least at the particular $k_z$ value corresponding to h$\nu$ = 21.218 eV,  in agreement with a previous report on A$_x$Fe$_2$Se$_2$ (A = K, Cs) \cite{Y_Zhang_arxiv2010}. This gap size leads to a $2\Delta/k_BT_c$ ratio of $\sim$ 7, indicating that the SC pairing in this material is in the strong coupling regime. The temperature dependence of spectra recorded at a single $k_F$ point are displayed in Fig. \ref{Fig_gap}(d), and the corresponding symmetrized EDCs are shown in Fig. \ref{Fig_gap}(e). Interestingly, the temperature ($T$) dependence of the SC gap given in Fig. \ref{Fig_gap}(d) indicates that the SC gap size may not close with temperature increasing, but rather fill in gradually up to $T_c$. Such behaviour has been reported recently in NaFe$_{0.95}$Co$_{0.05}$As \cite{ZH_Liu2010}.

\begin{figure}
\includegraphics[width=8.6cm]{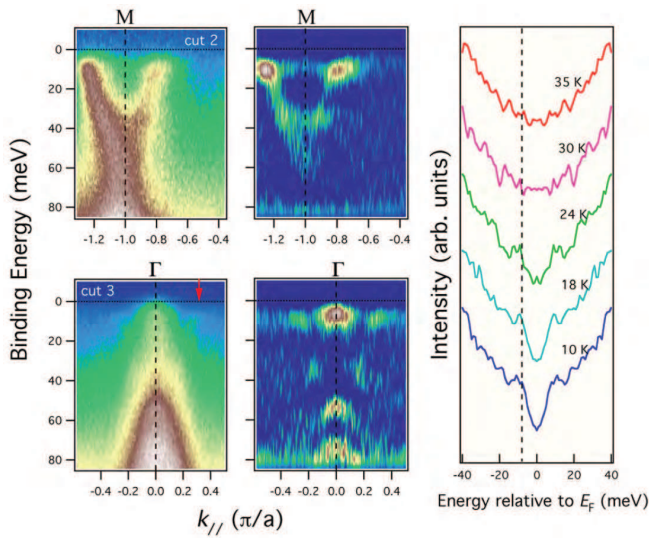}
\caption{\label{Fig_structure}(Colour on-line) (a) ARPES intensity plot along a cut passing through the M point [cut3 from Fig. \ref{Fig_map}(a)]. (b) Corresponding energy-second derivative intensity plot. (c) ARPES intensity plot along a cut passing through the $\Gamma$ point [cut3 from Fig. \ref{Fig_map}(a)]. (d) Corresponding energy-second derivative intensity plot. (e) Temperature evolution of the symmetrized EDCs at the $k_F$ position indicated by an arrow in (c). }
\end{figure}

The FS pockets at the M point encloses an area corresponding to 4.5 \% of the Brillouin zone (1 Fe site/unit cell description). This leads to a much smaller electron doping than expected from the Luttinger theorem. However, previous ARPES measurements on other iron-based superconductors indicate that there are two electronlike FS pockets with similar $k_F$'s centred at the M point \cite{Ding_EPL, Nakayama_EPL2009, Terashima_PNAS2009, Sekiba_NJP2009, ZH_Liu2010}. To confirm that this observation holds for Tl$_{0.63}$K$_{0.37}$Fe$_{1.78}$Se$_2$, we display the ARPES intensity plot along cut2 [see Fig. \ref{Fig_map}(a)] in Fig. \ref{Fig_structure}(a), and the corresponding energy-second derivative intensity plot in Fig. \ref{Fig_structure}(b). The data suggest that there are two distinct bands with bottoms around 40 and 60 meV, respectively. Within our experimental resolution, the bands have approximately the same $k_F$ values. 

The approximate 2-fold degeneracy of the electronlike FS pocket at M leads to an electron counting of 18\% electron per Fe, which is still not enough to afford for the 31.5\% electron per Fe expected from the Luttinger theorem for this material. We now turn our attention to the band structure around the $\Gamma$ point. Fig. \ref{Fig_structure}(c) and Fig. \ref{Fig_structure}(d) give the ARPES intensity plot along cut3 [see Fig. \ref{Fig_map}(a)] and the corresponding second derivative intensity plot along energy, respectively. The strongest feature is a holelike band topping around 50 meV below $E_F$. We also found a tiny electronlike feature with a bottom around 10 meV below $E_F$ that has been reported previously in A$_x$Fe$_2$Se$_2$ (A = K, Cs) \cite{Y_Zhang_arxiv2010}. In addition, we observe a large electronlike band at $\Gamma$ with a $k_F$ similar to that of the electron bands at M. The inclusion of the electronlike bands at the $\Gamma$ point raises the electron counting to 32\% electrons per Fe, which is in reasonable agreement with the nominal composition. As shown in Fig. \ref{Fig_structure} (e), where the EDC spectra at $k_F$ are presented as a function of temperature, a SC gap that vanishes above $T_c$ is found along the large electronlike FS at $\Gamma$. Interestingly, the SC gap size ($\sim$ 8 meV) is almost the same as the one we find at the M point. 

\begin{figure}
\includegraphics[width=8.6cm]{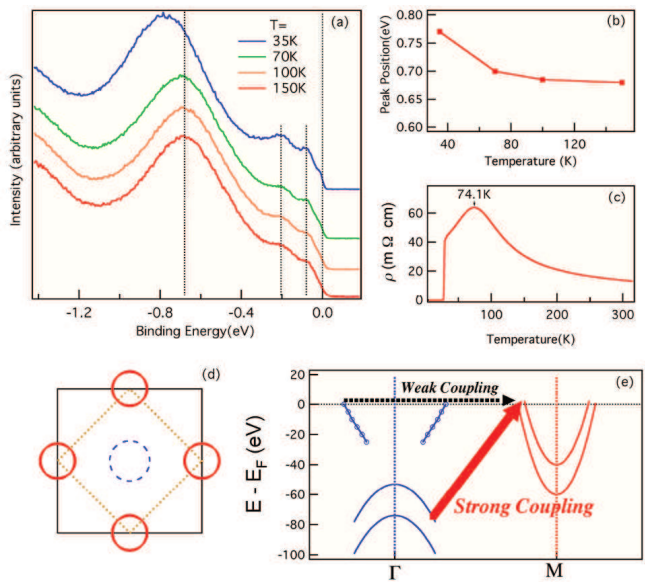}
\caption{\label{Fig_final}(Colour on-line) (a) Temperature evolution of the wide energy range spectrum at the $\Gamma$ point. (b) Temperature dependence of the peak position of the broad incoherent feature around 0.8 eV. (c) Temperature dependence of the resistivity of Tl$_{0.63}$K$_{0.37}$Fe$_{1.78}$Se$_2$, which indicates a metal-insulator transition at 74.1 K. (d) Schematic FS of Tl$_{0.63}$K$_{0.37}$Fe$_{1.78}$Se$_2$. (e) Schematic band structure of Tl$_{0.63}$K$_{0.37}$Fe$_{1.78}$Se$_2$.}
\end{figure}

Another major difference we observed in $K_{0.8}$Fe$_{1.7}$Se$_2$ from other iron-based superconductors is the presence of a large incoherent peak at $\sim$ 0.8 eV \cite{Qian_KFeSe2010}. This incoherent peak is also observed in Tl$_{0.63}$K$_{0.37}$Fe$_{1.78}$Se$_2$, as shown in Fig. \ref{Fig_final}(a). While this broad peak is found to be dispersionless in $k$-space, it has a drastic temperature dependence between 35 K and 150 K, as shown in Fig. \ref{Fig_final}(a). The energy shift at this temperature range is about 100 meV, with much of the shift occurring below 100 K, as indicated in Fig. \ref{Fig_final}(b). This shift corresponds well with the metal-nonmetal crossover around 74 K observed in the resistivity of this material, which is shown in Fig. \ref{Fig_final}(c). Interestingly, the low-energy valence band below 0.3 eV is relatively insensitive to this crossover. This unusual dichotomy of energy shift between the high-energy incoherent feature and the low-energy dispersion could be consistent with the doped Mott insulator picture where quasiparticles emerge from a gapped incoherent background. These quasiparticles with renormalized effective mass and coherence residue form the observed low-energy dispersion and FS displayed in Figs. \ref{Fig_final}(d). Strong pairing observed on these FSs may be affected by the interaction happening beyond the vicinity of $E_F$ [see Fig. \ref{Fig_final}(e)], even at an energy scale as large as the onsite Coulomb interactions.

The electronic structure of Tl$_{0.63}$K$_{0.37}$Fe$_{1.78}$Se$_2$ imposes severe constraints on the possible SC pairing mechanisms. The observation of nearly isotropic SC gap along two electron FS sheets near M is more consistent with scenarios for which the SC gap has the same sign on these two FSs. If they have opposite signs as for the $d$-wave gap would predict, nodes would likely emerge when the two bands hybridize. More significantly, superconductivity at high-temperature without holelike FS pocket strongly weakens the scenario of quasi-nesting-induced pairing of itinerant carriers.  It is known that only the nested electron-hole FS pockets would lead to a logarithmic divergence in the particle-hole scattering channel, although interband scattering between two electronlike FS pockets as a key ingredient cannot be ruled out. On the other hand, it is possible that particle-hole scattering can utilize the``sinking" hole pocket which is not far from the Fermi energy \cite{DH_Lee2011}. In this case, a strong coupling with local pairing may be favored. In fact, good agreement with SC gap functions derived from a local picture were already found for other iron-pnictides \cite{Nakayama_EPL2009,ZH_Liu2010, Wray_PRB2008, Nakayama_PRB2011, YM_Xu_NPhys2011}.

In conclusion, our high-resolution ARPES measurements of the highly electron-doped Tl$_{0.63}$K$_{0.37}$Fe$_{1.78}$Se$_2$ superconductor reveal nearly isotropic superconducting gaps on the two nearly degenerated electron FS sheets at the M point. The SC gap with an amplitude of $\sim$ 8.5 meV and closes above $T_c$, resulting in a pairing strength (2$\Delta$/$k_BT_c$ of $\sim$ 7) twice stronger than the weak-coupling BCS value. In addition, an unexpected electron Fermi surface with similar $k_F$ and $\Delta$ is observed around the zone centre, along with a holelike band sinking $\sim$ 50 meV below the Fermi energy. On the larger energy scale, the dispersionless incoherent peak at 0.7 - 0.8 eV shows a significant energy shift of $\sim$ 100 meV when going through the metal-nonmetal crossover around 74 K, while the low-energy band structure is relatively unaffected.

% eventual AF $(\pi,0)$ or ferromagnetic $(\pi,\pi)$ instabilities in this case are expected to be weaker and less robust upon doping. 

%\begin{acknowledgments}
We acknowledge W.-C. Jin, W.-D. Kong and H. Miao for technical assistance and X. Dai, Z. Fang, D.-H. Lee and D. J. Scalapino for useful discussions. This work is supported by Chinese Academy of Sciences (grant No. 2010Y1JB6), National Basic Research (973) Program of China (grants No. 2010CB923000, No. 2011CBA0010), and Nature Science Foundation of China (grants No. 10974175, No. 11004232, and No. 11050110422).
%\end{acknowledgments}

\bibliography{biblio_en}

\end{document}